\documentclass[aps,prd,twocolumn,superscriptaddress,showpacs,amsmath,amssymb,nofootinbib,preprintnumbers]{revtex4-1}
\usepackage{graphicx}
\usepackage{epstopdf}
\usepackage{float}
\usepackage{color}
\usepackage[toc,page]{appendix}
\usepackage[usenames,dvipsnames]{xcolor}
\usepackage{tensor}
\usepackage[normalem]{ulem}

\newcommand{\be}{\begin{equation}}
\newcommand{\ee}{\end{equation}}
\newcommand{\ba}{\begin{eqnarray}}
\newcommand{\ea}{\end{eqnarray}}

\newcommand{\beq}{\begin{equation}}
\newcommand{\eeq}{\end{equation}}
\newcommand{\beqa}{\begin{eqnarray}}
\newcommand{\eeqa}{\end{eqnarray}}

\begin{document}
\title{Rotating Gauss--Bonnet BTZ Black Holes}

\author{Robie A. Hennigar}
\email{rhennigar@mun.ca}
\affiliation{Department of Mathematics and Statistics, Memorial University of Newfoundland, St. John's, Newfoundland and Labrador, A1C 5S7, Canada }

\author{David Kubiz\v n\'ak}
\email{dkubiznak@perimeterinstitute.ca}
\affiliation{Perimeter Institute, 31 Caroline St. N., Waterloo,
Ontario, N2L 2Y5, Canada}
\affiliation{Department of Physics and Astronomy, University of Waterloo,
Waterloo, Ontario, Canada, N2L 3G1}

\author{Robert B. Mann}
\email{rbmann@uwaterloo.ca}
\affiliation{Department of Physics and Astronomy, University of Waterloo,
Waterloo, Ontario, Canada, N2L 3G1}
\affiliation{Perimeter Institute, 31 Caroline St. N., Waterloo,
Ontario, N2L 2Y5, Canada}

\date{May 27, 2020}

\begin{abstract}
We obtain rotating black hole solutions to the novel 3D Gauss--Bonnet theory of gravity recently proposed.
These solutions generalize the BTZ metric and are not of constant curvature.  They possess an ergoregion and outer horizon, but do not have an inner horizon.  We present their basic properties and show that they break the universality of thermodynamics present for their static charged counterparts, whose properties we also discuss.
Extending our considerations to higher dimensions, we also obtain novel 4D Gauss--Bonnet rotating black strings.
\end{abstract}

\maketitle

\section{Introduction}

Higher-curvature theories of gravity continue to remain of considerable interest, in part because most approaches to quantum gravity suggest that the Einstein--Hilbert action is modified by such corrections, and in part because such theories provide a new arena for testing our understanding of classical gravity in strong gravitational fields.  The most promising and widely studied candidates are Lovelock theories~\cite{Lovelock:1971yv}: they are the most general theories  built from the Riemann curvature tensor that retain second-order equations of motion for the metric. However they have the property that their additional contributions to the action are either topological or identically zero for $D < 5$.

A new proposal~\cite{Glavan:2019inb} for circumventing this limitation for Gauss--Bonnet gravity (the simplest of the Lovelock theories)  has generated considerable interest recently.  By treating the spacetime dimension as a parameter of the theory and rescaling the Gauss--Bonnet coupling $\alpha$, it is possible to obtain $D=4$ and $D=3$ versions of this theory.   Although the original proposal involved taking limits of solutions to the field equations, raising a variety of issues of consistency~\cite{Gurses:2020ofy, Hennigar:2020lsl, Bonifacio:2020vbk}, it is possible to obtain consistent actions in $D<5$ without making any assumptions about either actual solutions or  extra dimensions~\cite{Fernandes:2020nbq, Hennigar:2020lsl, Hennigar:2020fkv}.  This approach is a generalization of one applied quite some time ago to obtain a $D\to 2$ limit of general relativity~\cite{Mann:1992ar}, and is consistent with a dimensional reduction procedure recently expounded~\cite{Lu:2020iav,Kobayashi:2020wqy}, provided the internal space is flat.
The resultant theory is the following scalar-tensor theory of gravity
\ba\label{SD}
S&=&\int d^D x \sqrt{-g}\Bigl[R-2\Lambda+\alpha\Bigl(\phi {\cal G}+4 G^{ab}\partial_a \phi \partial_b \phi\nonumber\\
&&\qquad\qquad\qquad -4(\partial \phi)^2 \Box \phi+2((\nabla\phi)^2)^2 \Bigr) \Bigr]\,,
\ea
where ${\cal G}=R_{abcd}R^{abcd}-4 R_{ab}R^{ab}+R^2$ is the Gauss--Bonnet term (which identically vanishes in $D<4$ dimensions) and $G_{ab}$ is the Einstein tensor. For $D=3$ there is no scalar propagating degree of freedom \cite{Lu:2020mjp}.

Much effort has been expended analyzing the consequences of lower-dimensional Gauss--Bonnet gravity, including
black holes,  star-like solutions, radiating and collapsing solutions,  cosmological solutions, black hole thermodynamics   and various physical implications (see citations of \cite{Glavan:2019inb}).   All black hole solutions so far have been for spherically symmetric metrics.

We present here the first example of an exact rotating solution in $D=3$ Gauss--Bonnet gravity \eqref{SD}.
Unlike previous solutions that were simply an embedding of the rotating Banados--Teitelboim--Zanelli (BTZ)~\cite{Banados:1992wn} metric that required curvature in the internal space  \cite{Ma:2020ufk}, our solutions are a novel generalization of the rotating BTZ metric and provide us with a new class of rotating black holes in $(2+1)$ dimensions without constant curvature that might be expected from quantum gravitational corrections to the classical action. Having such solutions will be of considerable value, since many problems intractable in higher-dimensions can be solved (or at least ameliorated) by considering BTZ black holes~\cite{Ross:1992ba, Carlip:1995qv, Cardoso:2001hn, Konoplya:2004ik}.

\section{Static black holes}

The field equations of the $D=3$ version of Gauss--Bonnet gravity \eqref{SD} yield the following solution \cite{Hennigar:2020fkv} (see also \cite{Konoplya:2020ibi, Ma:2020ufk}):
\be\label{ds23d}
ds^2=-f_{GB}dt^2+\frac{dr^2}{f_{GB}}+r^2 d\varphi^2\,,\quad
\phi = \ln(r/l)\,,
\ee
where
\be\label{formula}
f^\pm_{GB}=-\frac{r^2}{2\alpha}\Bigl(1 \pm \sqrt{1+\frac{4\alpha}{r^2}f_{E}}\Bigr)\,,
\ee
is the Gauss--Bonnet generalization of the (static) Einstein theory BTZ metric function \cite{Banados:1992wn}
\be\label{fE}
f_E=\frac{r^2}{\ell^2}  - m\,,
\ee
with $m$ and $l$ integration constants and $\Lambda=-1/\ell^2$.

Denoting the black hole solution by $f_{GB} \equiv f^{-}_{GB}$, it is simple to show that not only are the horizons located at the same value of $r_+$ for the two theories, but  the temperatures
\be
T=\frac{f'_E(r_+)}{4\pi}=\frac{f'_{GB}(r_+)}{4\pi}
\ee
are also the same. Futhermore,  since the scalar field $\phi$ is finite on the horizon, the entropies coincide
\be
S=\frac{\pi r_+}{2}
\ee
despite the fact that the Gauss--Bonnet theory \eqref{SD} is a higher-curvature theory and the entropy is calculated using the Wald prescription \cite{Iyer:1994ys}.

The result is a `universal thermodynamics':
\ba
M&=&\frac{m}{8}=\frac{r_+^2}{8\ell^2}\,,\quad T=\frac{f'}{4\pi}=\frac{r_+}{2\pi \ell^2}\,,\quad S=\frac{\pi r_+}{2}\,,\nonumber\\
P&=&\frac{1}{8\pi \ell^2}\,,\quad V=\pi r_+^2\,,\quad \psi_\alpha=0\,,
\ea
identical to that of the familiar BTZ black hole in Einstein gravity, albeit with the additional potential $\psi_\alpha$ conjugate to $\alpha$.  This is quite intriguing -- even though the curvature is not constant, the thermodynamic parameters are the same for any value of $\alpha$. We also obtain
\be
\delta M=T\delta S+V\delta P+\psi_\alpha \delta \alpha\,,\quad 0=TS-2PV+2\psi_\alpha \alpha\,,
\ee
which are the standard first law and Smarr relations \cite{Kastor:2010gq}.
It is known that the thermodynamic properties of higher-dimensional Lovelock black branes are identical to those of black branes in Einstein gravity~\cite{Cadoni:2016hhd,Hennigar:2017umz}. The observations here are consistent with this property,  now extended to lower dimensions.

Charged generalizations of the solution \eqref{ds23d} in the Maxwell and Born--Infeld theories are discussed in the appendix.

\section{Rotating black holes}

\subsection{BTZ solution}
Before proceeding to consider the rotating Gauss--Bonnet BTZ black hole, it will be helpful to make some remarks on an alternate form for the rotating BTZ metric in Einstein gravity. This metric can be obtained by `boosting' the static BTZ black hole according to \cite{Lemos:1994xp, Lemos:1995cm}:
\be\label{tmf}
t \to \Xi t - a \varphi \, , \quad \varphi \to  \frac{at}{\ell^2} - \Xi \varphi \, ,\quad \Xi^2 = 1 + \frac{a^2}{\ell^2} \, ,
\ee
yielding the metric
\be\label{btzfunny}
ds^2=-f_E(\Xi dt -a d\varphi)^2+\frac{r^2}{\ell^4}(adt-\Xi \ell^2 d\varphi)^2+\frac{dr^2}{f_E}\,,
\ee
where $f_E$ is the metric function \eqref{fE}. The usual form of the metric
\ba\label{standard}
ds^2&=&-\hat f dt^2+\hat{r}^2\Bigl(d\varphi -\frac{J}{2\hat{r}^2} dt\Bigr)^2+\frac{d\hat{r}^2}{\hat f}\,,\nonumber\\
\hat f &=& \frac{\hat{r}^2}{\ell^2}-\hat m +\frac{J^2}{4\hat{r}^2}\,,
\ea
for the rotating BTZ black hole can be recovered by a simple shift of the radial coordinate:
\be
 {\hat r}^2=-f_E a^2+r^2\Xi^2\quad \Rightarrow \quad
r = \sqrt{\hat{r}^2-ma^2}\,,
\ee
together with setting
\be
J=2\Xi am\,,\quad \hat m=m\Bigl(1+\frac{2a^2}{\ell^2}\Bigr)\,.
\ee

Written in the `Kerr-like' coordinates~\eqref{btzfunny}, the event horizon is located at the largest root of $f_E = 0$, while the inner horizon is located at $r = 0$. The corresponding thermodynamic quantities are
\ba
M&=&\frac{m(2\Xi^2-1)}{8}\,,\quad  J=\frac{1}{4}\Xi am\,,\quad \Omega=\frac{a}{\Xi \ell^2}\,,\nonumber\\
T&=&\frac{f_E'}{4\pi \Xi}=\frac{r_+}{2\pi\Xi \ell^2}\,,\quad S=\frac{\pi \Xi r_+}{2}\,,
\nonumber\\
P&=&\frac{1}{8\pi \ell^2}\,,\quad V=\pi \Xi^2 r_+^2\,,
\ea
and obey the standard first law and Smarr relations:
\be
\delta M=T\delta S+\Omega \delta J+V\delta P\,, \quad
0 = TS-2PV+\Omega J\,.
\ee

While there is no clear advantage to using~\eqref{btzfunny} in Einstein gravity, we will now see that this form of the metric allows for considerable simplification when obtaining solutions with rotation to the equations of  \eqref{SD}.

\subsection{Gauss--Bonnet black holes}

To obtain a rotating solution in the Gauss--Bonnet theory we perform the same type of boost used in the Einstein case, but modified to account for the fact that the higher-curvature corrections result in modifications to the AdS length scale of the theory:
\be\label{boost}
t\to \Xi_{\rm eff} t -a \varphi\,, \  \varphi \to \frac{a t}{\ell_{\rm eff}^2} -\Xi_{\rm eff} \varphi\,,  \  \Xi_{\rm eff}^2 = 1 + \frac{a^2}{\ell_{\rm eff}^2}\,,
\ee
applied to the metric \eqref{ds23d}, where
\be\label{Xeff}
\ell_{\rm eff}= \sqrt{ \frac{2 \alpha}{\sqrt{X_\alpha}-1}}\,,\quad X_\alpha = 1+\frac{4\alpha}{\ell^2}\,.
\ee
This yields
\ba
ds^2&=&\!-f_{GB}(\Xi_{\mbox{\tiny eff}} dt \!-\!a d\varphi)^2\!+\!\frac{r^2}{\ell_{\mbox{\tiny eff}}^4}(adt\!-\!\Xi_{\mbox{\tiny eff}} \ell_{\mbox{\tiny eff}}^2 d\varphi)^2 \!+\!\frac{dr^2}{f_{GB}}\,,\nonumber\\
\phi& =& \ln(r/l)\,,
\label{rotbtz}
\ea
with $f_{GB}$ given by $f^{-}_{GB}$ in \eqref{formula}, and $l$ an arbitrary integration constant.
Had we instead used the `bare' cosmological length scale $\ell$ in the boost, the resulting metric would rotate at infinity, complicating the thermodynamics \cite{Gibbons:2004ai}.

The metric  \eqref{rotbtz} is an exact solution to the field equations of \eqref{SD} and is the Gauss--Bonnet generalization of the rotating BTZ black hole.  However, there are some important differences worth commenting on.
Consider constant $r$-surfaces in the geometry. The determinant of the induced metric on these surfaces is
\be
\sigma = -r^2 f_{GB} \, .
\ee
The surfaces corresponding to $r= 0$ and $f_{GB} = 0$ are null and are horizons in this coordinate system.  For the  ordinary BTZ black hole \eqref{btzfunny}, the largest root of 
$f_E=0$
corresponds to the event horizon while $r = 0$ corresponds to the inner Cauchy horizon. However, in the Gauss--Bonnet case, the metric function need not even extend all the way to $r = 0$. For positive coupling the metric function is real only when~\cite{Hennigar:2020fkv}
\be
r >  2 \sqrt{\frac{\alpha m}{X_\alpha}} ,
\ee
and at this value of $r$ there is a branch singularity. Thus, it is clear that when $\alpha >   0$ the metric cannot extend to $r = 0$. At the branch singularity, the metric (monotonically) approaches
\be
f_{GB} \to -\frac{2 m }{X_\alpha} \, ,
\ee
and so crosses zero only once. In this case, a curvature singularity is reached prior to the location of the would-be inner horizon.

If $\alpha < 0$ the situation is more interesting, since the metric extends all the way to $r = 0$. However, in the vicinity of $r = 0$ the metric function behaves as
\be
f(r) \to \sqrt{-\frac{m}{\alpha}} r + {\cal O}(r^2)
\ee
and there is a curvature singularity at the origin~\cite{Hennigar:2020fkv}.

Despite this exotic causal structure, the rotating Gauss--Bonnet BTZ black hole possesses other features expected of a rotating geometry. Ergoregions are regions of spacetime where the Killing field $\partial/\partial t$ becomes spacelike. In the present case, this corresponds to the following condition:
\be
g_{tt} = -f_{GB} \Xi_{\rm eff}^2 + \frac{r^2 a^2}{\ell_{\rm eff}^2} > 0 \, .
\ee
Clearly, at the horizon (and necessarily also for some region outside of it) $g_{tt} > 0$ and an ergoregion is  present.

\begin{figure}
\centering
\includegraphics[width = 0.45\textwidth]{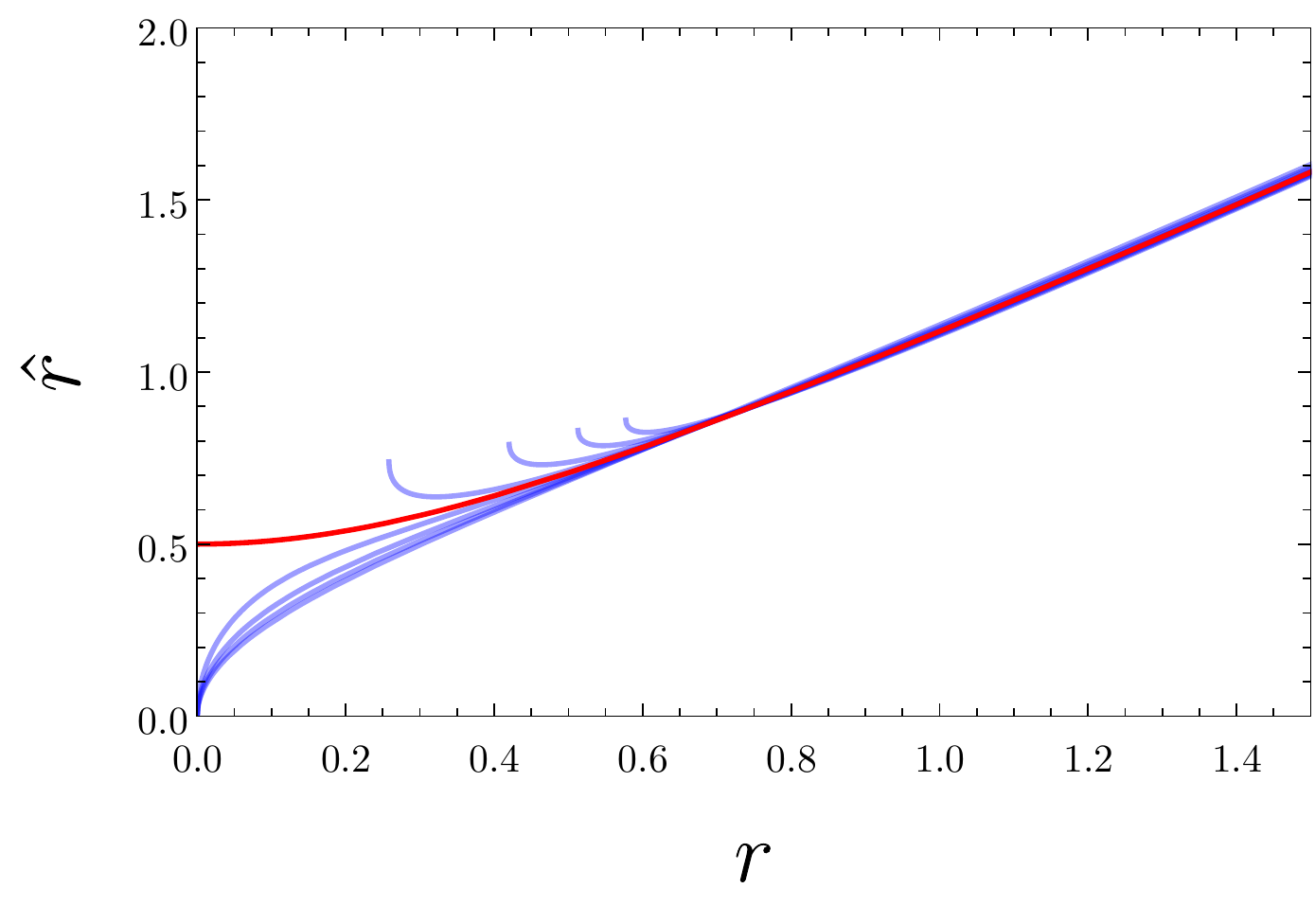}
\caption{Here we show plots of $\hat{r}$ vs.~$r$ for the BTZ black hole (red), and the Gauss--Bonnet corrected BTZ black hole (blue curves). The blue curves lying above the red curve have $\alpha  > 0$, while those lying below the red curve have $\alpha < 0$. In this plot we have fixed $m = 1, a  = 1/2, \ell = 1$. The blues curves (from bottom to top) correspond to $\alpha = -1/8$ to $\alpha = +1/8$ in increments of $1/28$.  At large-$r$, all cases have $\hat{r} \approx r$.}
\label{F1}
\end{figure}

The metric can be cast into `ADM form', which is commonly used for the rotating BTZ black hole~\eqref{standard}:
\be
ds^2 = - N^2 dt^2 + \hat{r}^2 \left(N^\varphi dt + d\varphi\right)^2 + \frac{d\hat{r}^2}{\hat{f}} \, .
\ee
This requires transforming the radial coordinate according to
\be\label{newR}
\hat{r}^2 = -f_{GB} a^2 + r^2 \Xi_{\rm eff}^2 = r^2 + \frac{2 a^2 m}{\sqrt{X_\alpha}+\sqrt{X_\alpha-\frac{4\alpha m}{r^2}}}
\, ,
\ee
with the lapse and shift given by
\be
N^2 = \frac{r^2 f_{GB}}{\hat{r}^2} \,, \quad N^\varphi = \frac{a \Xi_{\rm eff}}{\hat{r}^2} \left[ f_{GB} - \frac{r^2}{\ell_{\rm eff}^2} \right]\, ,
\ee
and analog of the metric function $\hat{f} = 1/g_{\hat{r}\hat{r}}$ is
\be
\hat{f}_{GB} = \frac{f_{GB} \left[2 r \Xi^2_{\rm eff} - a^2 f'_{GB} \right]^2}{4 \hat{r}^2} \, .
\ee

Inverting~\eqref{newR} to obtain $r(\hat{r})$ is algebraically straightforward, but the resultant expressions are quite cumbersome, and so we  plot $\hat{r}$ as a function of $r$ in Fig.~\ref{F1}. For $\alpha > 0$, we see that $\hat{r}$ fails to exist beyond the branch singularity (where the spacetime curvature also diverges). For $\alpha < 0$, $\hat{r}\to 0$  as $r \to 0$. This is a consequence of the fact mentioned above that $f_{GB} \to 0$ as $r \to 0$. As such, the behaviour of this coordinate is markedly from the ordinary BTZ black hole.

\begin{figure}
\centering
\includegraphics[width = 0.45\textwidth]{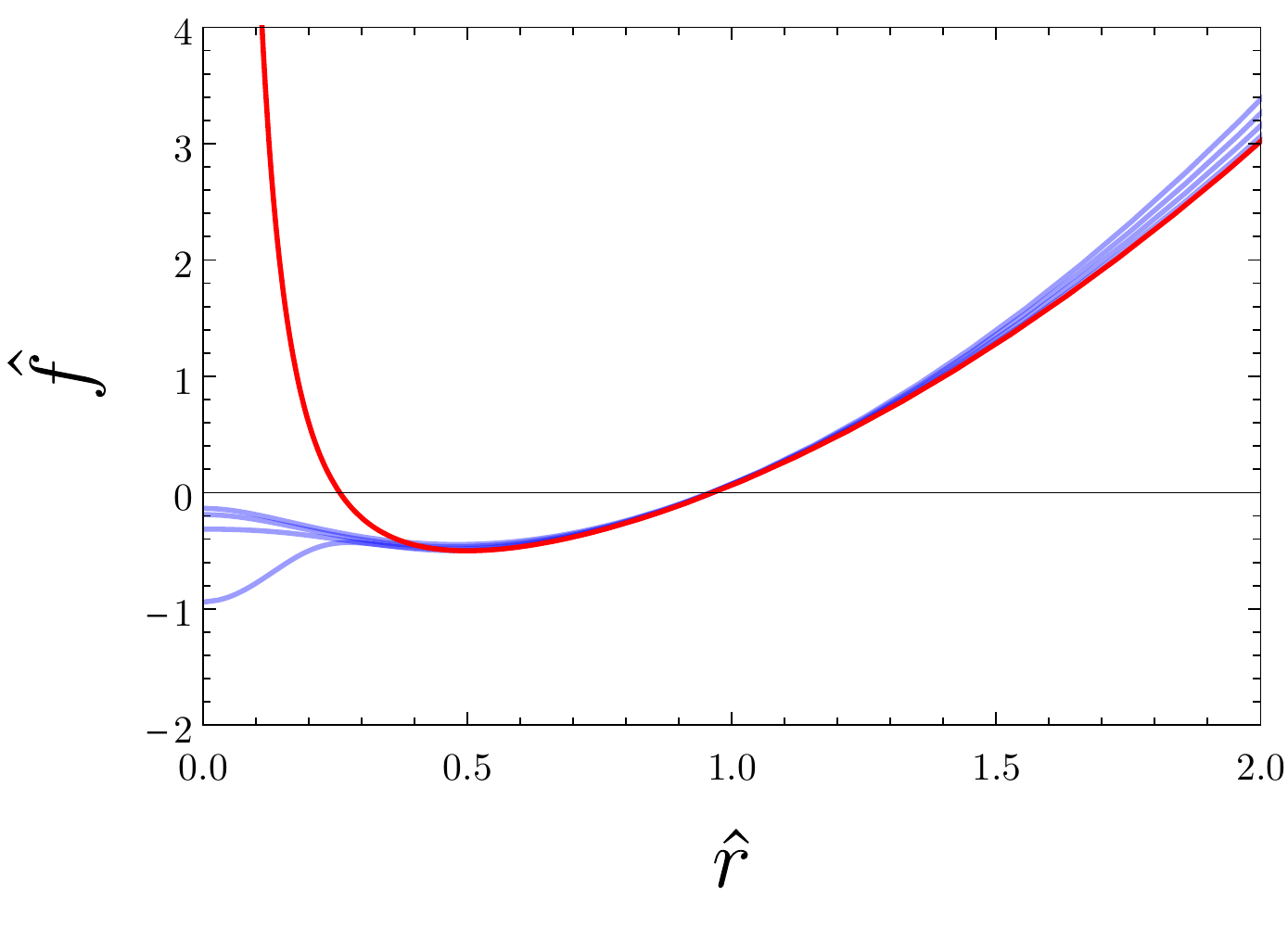}
\caption{Here we show plots of $\hat{f}$ vs.~$\hat{r}$ for the BTZ black hole (red), and the Gauss--Bonnet corrected BTZ black hole (blue curves) with $\alpha < 0$. The curves correspond to $\alpha = -1/8$ to $0$ in increments of $1/28$ (top to bottom). We have set the physical parameters $(M, J) = (1, 1/2)$ for all cases, with $\ell = 1$. At large distances, the metric function behaves as $\hat{f}_{GB} \sim \hat{r}^2/\ell_{\rm eff}^2$.}
\label{fPlot}
\end{figure}

We plot the metric function $\hat{f}$ in Fig.~\ref{fPlot} instead of presenting the cumbersome expression, with  the BTZ case in red.
At large distances the metric function behaves as $\hat{f}_{GB} \sim \hat{r}^2/\ell_{\rm eff}^2$, while near the event  horizon the curves are remarkably close to one another. Significant differences appear deep inside the black hole, with the Gauss--Bonnet-BTZ solution failing to exhibit an inner horizon. The behaviour of the lapse is qualitatively the same as $\hat{f}$, with the only notable difference being that $N \to 0$ as $\hat{r} \to 0$ for all negative coupling.

We find that the thermodynamic parameters are now
\ba
M&=&\frac{m(2\Xi_{\mbox{\tiny eff}}^2-1)}{8}\,,\quad  J=\frac{\Xi_{\mbox{\tiny eff}} am}{4}\,,\quad \Omega=\frac{a}{\Xi_{\mbox{\tiny eff}} \ell^2_{\mbox{\tiny eff}}}\,,\nonumber\\
T&=&\frac{f_{GB}'}{4\pi \Xi_{\mbox{\tiny eff}}}=\frac{r_+}{2\pi\ell^2\Xi_{\mbox{\tiny eff}}}\,,\quad S=\frac{\pi \Xi_{\mbox{\tiny eff}} r_+}{2}\,,
\nonumber\\
P&=&\frac{1}{8\pi \ell^2}\,,\quad V= \pi r_+^2 \Bigl( 1+ \frac{a^2}{\ell^2 \sqrt{X_\alpha}}\Bigr)\,,\nonumber\\
\psi_\alpha&=&  \frac{r_+^2 a^2}{16\alpha^2 \ell^2 \sqrt{X_\alpha}}\Bigl(\sqrt{X_\alpha}-1-\frac{2\alpha}{\ell^2}\Bigr)\,,
\ea
and obey the respective standard first law and Smarr relation
\ba
\delta M&=&T\delta S +\Omega \delta J +V\delta P+\psi_\alpha \delta \alpha\,,\label{first}\\
0&=&TS-2PV + \Omega J+2\psi_\alpha \alpha\,.\label{Smarr}
\ea
The thermodynamics of this rotating black hole
is significantly different from that of its BTZ counterpart: universality of thermodynamics is no longer valid in the presence of rotation. One can easily check that when $\alpha>0$  the isoperimetric ratio \cite{Cvetic:2010jb}  is
${\cal R}=1-\frac{a^2}{2\ell^2}\frac{\alpha}{\ell^2+a^2}+O(\alpha^2)$, indicating that these rotating black holes are super-entropic \cite{Hennigar:2014cfa}.

\section{Rotating black string}

It is straightforward to generalize our considerations to $D=4$.  We find the following generalization of the black string \cite{Lemos:1994xp, Lemos:1995cm}:
\ba
ds^2&=&-f_{GB}dt^2+\frac{dr^2}{f_{GB}}+r^2 d\varphi^2+r^2 dx^2\,,\quad \phi=\ln(r/l)\,,\nonumber\\
f_{GB}&=&\frac{r^2}{2\alpha}\Bigl(1\pm \sqrt{1-\frac{4\alpha}{r^2} f_E}\Bigr)\,, \quad f_E=\frac{r^2}{\ell^2}-\frac{2M}{r}\,,
\ea
and its rotating generalization
\ba
ds^2&=&-f_{GB}(\Xi_{\mbox{\tiny eff}} dt -a d\varphi)^2+\frac{r^2}{\ell_{\mbox{\tiny eff}}^4}(adt-\Xi_{\mbox{\tiny eff}} \ell_{\mbox{\tiny eff}}^2 d\varphi)^2\nonumber\\
&&+\frac{dr^2}{f_{GB}}+r^2 dx^2\,,
\ea
where
\be
\Xi_{\mbox{\tiny eff}}=\sqrt{1+\frac{a^2}{\ell_{\mbox{\tiny eff}}^2}}\,,\quad  \ell_{\mbox{\tiny eff}}=\sqrt{\frac{2\alpha}{1-\sqrt{X_{-\alpha}}}} \,,
\ee
and the sign difference in $\alpha$ compared to  \eqref{Xeff}  is a consequence of setting $D=4$ in \eqref{SD}.

The thermodynamic quantities (per unit length of the string) are
\ba
M&=&\frac{m\Xi_{\mbox{\tiny eff}}(\Xi_{\mbox{\tiny eff}}^2+1)}{4}\,,\quad  J=\frac{am\Xi_{\mbox{\tiny eff}}^2}{4}\,,\quad \Omega=\frac{a}{\Xi_{\mbox{\tiny eff}} \ell^2_{\mbox{\tiny eff}}}\,,\nonumber\\
T&=&\frac{f_{GB}'}{4\pi \Xi_{\mbox{\tiny eff}}}=\frac{3}{4}\frac{r_+}{\pi \ell^2\Xi_{\mbox{\tiny eff}}}\,,\quad S=\frac{\pi \Xi_{\mbox{\tiny eff}} r_+^2}{2}\,, \quad P = \frac{1}{8\pi \ell^2}\,,
\nonumber\\
\psi_\alpha&=& \frac{a^2 r_+^3}{4\ell^6 (1+\sqrt{X_{-\alpha}} )^2 \sqrt{X_{-\alpha}}} \sqrt{1\!+\! \frac{ 2 a^2 }{\ell^2(1+ \sqrt{X_{-\alpha}}) }}\,,\nonumber\\
V &=& \frac{\pi r_+^3 (a^2 + 4\sqrt{X_{-\alpha}} \ell^2)}{6\ell^2 \sqrt{X_{-\alpha}}} \sqrt{1\!+\! \frac{ 2 a^2 }{\ell^2(1+ \sqrt{X_{-\alpha}}) }}\,,
\ea
and obey the standard first law and (4D) Smarr relations:
\ba
\delta M&=&T\delta S+\Omega \delta J+V\delta P+\psi_\alpha \delta \alpha\,,\nonumber\\
M&=&2(TS-PV+\Omega J+\psi_\alpha \alpha)\,.
\ea

We expect that a similar construction works for black branes if we set $D>4$ in \eqref{SD}  as well, generalizing thus the construction in \cite{Awad:2002cz, Dehghani:2002wn, Dehghani:2006dh}. Contrary to the statements in those paper, however, in order to formulate correct thermodynamics of these objects one needs to employ the effective AdS radius $\ell_{\mbox{\tiny eff}}$ rather than the bare AdS radius $\ell$ in the boost formulae. Once this is done a metric that is non-rotating at infinity is obtained
and its thermodynamics (with correct thermodynamic quantities) can be constructed.

\section{Conclusions}

The novel rotating Gauss-Bonnet BTZ black holes obtained in this paper have a number of remarkable properties.
As with their BTZ counterparts, they posses an ergoregion and an outer horizon. However they possess no inner horizon -- instead they have a cuvature singularity that precludes its formation.  If the $\alpha$-dependent terms in \eqref{SD} arise from quantum gravitational effects, this would be evidence that quantum gravity eliminates
unstable inner horizons.  Furthermore, rotation breaks the universality of the thermodynamics present in the spherically symmetric case.

We have also also shown that rotating black string solutions exist for $D=4$ Gauss-Bonnet gravity. Their horizon and ergosphere properties are qualitatively similar to the $D=3$ case, and we expect our construction extends to higher-dimensions.

It would be interesting to explore connections between this class of rotating metrics and strong cosmic censorship, which is now known for fail for the rotating BTZ black hole~\cite{Dias:2019ery,Balasubramanian:2019qwk}.  Such a study would provide insight as to the role non-linearity in curvature might play in this regard.

\section*{Acknowledgements}

This work was supported in part by the Natural Sciences and Engineering Research Council of Canada.
R.A.H.\ is supported by the Natural Sciences and Engineering Research Council of Canada
through the Banting Postdoctoral Fellowship program
D.K.\ acknowledges the Perimeter Institute for Theoretical Physics  for their support. Research at Perimeter Institute is supported in part by the Government of Canada through the Department of Innovation, Science and Economic Development Canada and by the Province of Ontario through the Ministry of Colleges and Universities.

\appendix

\section{Charged Black Holes}

The static solution \eqref{ds23d}--\eqref{fE} can be easily generalized to include the effects of the (possibly non-linear) electromagnetic  field, by adding  the following action:
\be\label{NLE}
S_{EM}=\int d^3 x \sqrt{-g}{\cal L}(F)\,,\quad F=F_{ab}F^{ab}\,
\ee
to the theory \eqref{SD}. Considering first the Maxwell theory
\be\label{Maxwell}
{\cal L}_M=-F\,,
\ee
we find that
\ba\label{chargedBH}
ds^2&=&-f_{GB}dt^2+\frac{dr^2}{f_{GB}}+r^2 d\varphi^2\,,\nonumber\\
f^\pm_{GB}&=&-\frac{r^2}{2\alpha}\Bigl(1\pm \sqrt{1+\frac{4\alpha}{r^2}f_E}\Bigr)\,,\quad
\phi =  \ln(r/l)\,,\quad
\label{MBTZ2}
\ea
describes for $f^{-}_{GB}$ a charged Gauss--Bonnet BTZ black hole,
provided the metric function $f_E$ and the vector potential $A$ correspond to those of the charged BTZ black hole in the Einstein gravity \cite{Banados:1992wn}:
\be
f_E=\frac{r^2}{\ell^2}-m-2e^2\ln(r/r_0)\,,\quad
A = -e\ln(r/r_0) dt\,,
\ee
with $r_0$, $l$, and $m$ arbitrary integration constants.

Similarly, for the Born--Infeld theory \cite{born1934foundations},
\be\label{BI}
{\cal L}_{BI}=-b^2\Bigl(\sqrt{1+\frac{2F}{b^2}}-1\Bigr)\,,
\ee
which reduces to the Maxwell Lagrangian \eqref{Maxwell}
in the limit $b\to \infty$, we find
that the solution takes the form  \eqref{MBTZ2}, where now
the Einstein metric function $f_E$ and vector potential $A$ are those of the
Born--Infeld charged BTZ black hole  \cite{Cataldo:1999wr, Myung:2008kd}:
\ba
f_E&=&\frac{r^2}{\ell^2}-m+\frac{1}{2}rb^2\Bigl(r-\sqrt{r^2+r_1^2}\Bigr)
\nonumber\\
&-&\frac{1}{2}r_0b^2\Bigl(r_0-\sqrt{r_0^2+r_1^2}\Bigr)-2e^2\ln \left(\frac{r+\sqrt{r^2+r_1^2}}{r_0+\sqrt{r_0^2+r_1^2}}\right)\,,\nonumber \\
A&=&-e\ln \left(\frac{r+\sqrt{r^2+r_1^2}}{r_0+\sqrt{r_0^2+r_1^2}}\right) dt\,,\quad r_1=\frac{2e}{b}\,,
\ea
where again $r_0, l, m$  are arbitrary integration constants.

We expect that the form   \eqref{MBTZ2} remains valid for charged Gauss--Bonnet BTZ black holes in any theory of non-linear electrodynamics characterized by the action \eqref{NLE}. If so, the charged Gauss--Bonnet BTZ black holes share the same horizon radius $r_+$ and the same temperature
\be
T=\frac{f'_E(r_+)}{4\pi}=\frac{f'_{GB}(r_+)}{4\pi}\,
\ee
with their Einstein gravity cousins. If  their vector fields also coincide, then the thermodynamics  is likewise
universal.

In particular, for the Born--Infeld theory we have the following thermodynamic quantities:
\ba
M&=&\frac{m}{8}\,,\quad T=\frac{f_E'(r_+)}{4\pi}=\frac{r_+}{2\pi \ell^2}\,,\quad S=\frac{1}{2}\pi r_+\,,\nonumber\\
Q&=&e\,,\quad \phi=-\frac{e}{2}\ln \left(\frac{r+\sqrt{r^2+r_1^2}}{r_0+\sqrt{r_0^2+r_1^2}}\right)\,,\nonumber\\
P&=&\frac{1}{8\pi \ell^2}\,,\quad V=\pi r_+^2\,,
\ea
valid both for the Einstein and Gauss--Bonnet theories. These quantities satisfy the following
first law of black hole thermodynamics:
\be
\delta M=T\delta S+\phi \delta Q+V\delta P\,,
\ee
where we have treated the integration constant $r_0$ to be a scale that is independent of the AdS radius $\ell$.\footnote{Alternatively, one could let the two coincide, which however, leads to exotic thermodynamics, see  \cite{Frassino:2015oca, Appels:2019vow} for a discussion.} To write the corresponding Smarr relation one would need to include the variations the cosmological constant, as well as the variations of $r_0$ and of the Born--Infeld parameter $b$ and include the corresponding `Born-Infeld polarization' term into considerations, see \cite{Gunasekaran:2012dq}. The Maxwell case  is recovered upon setting $b\to \infty$.

The obtained solutions in this appendix can be considered as $D=3$ versions of the charged Gauss--Bonnet black holes recently studied in four dimensions \cite{Fernandes:2020rpa, Yang:2020jno}.  We shall consider extending these solutions to include rotation in a future paper.


\providecommand{\href}[2]{#2}\begingroup\raggedright\endgroup

\end{document}